\documentclass[a4paper]{spie}  

\usepackage[]{graphicx}
\usepackage[]{abbrevs}

\title{An update on the SCUBA-2 project} 

\newcommand{\AmS}{{\protect\the\textfont2
  A\kern-.1667em\lower.5ex\hbox{M}\kern-.125emS}}
\def\sqr#1#2{{\vcenter{\vbox{\hrule height.#2pt
        \hbox{\vrule width.#2pt height#1pt \kern#1pt
          \vrule width.#2pt}
        \hrule height.#2pt}}}\relax}
\def\square{\mathchoice\sqr45\sqr45\sqr{3.1}4\sqr{2.5}4}

\author{Michael D. Audley\supit{a}, Wayne S. Holland\supit{a}, Trevor Hodson\supit{a}, Mike MacIntosh\supit{a}, Ian Robson\supit{a}, Kent Irwin\supit{b}, Gene Hilton\supit{b}, William Duncan\supit{b}, Carl Reintsema\supit{b}, Anthony Walton\supit{c}, William Parkes\supit{c}, Peter Ade\supit{d}, Ian Walker\supit{d}, Michel Fich\supit{e}, Jan Kycia\supit{e}, Mark Halpern\supit{f}, David A. Naylor\supit{g}, George Mitchell\supit{h}, and Pierre Bastien\supit{i} 
\skiplinehalf
\supit{a}UK Astronomy Technology Centre, Royal Observatory, Edinburgh, EH9 3HJ, UK; \\
\supit{b}National Institute of Standards and Technology, 325 Broadway, Boulder, CO 80305, USA; \\
\supit{c}The Scottish Microelectronics Centre, University of Edinburgh, Edinburgh, EH9 3JF, UK; \\
\supit{d}Cardiff University, Cardiff, CF24 3YB, UK; \\
\supit{e}University of Waterloo, Waterloo, Ontario N2L 3G1, Canada; \\
\supit{f} University of British Columbia
Vancouver, BC V6T 1Z1, Canada; \\
\supit{g}University of Lethbridge,
Lethbridge, Alberta
T1K 3M4, Canada; \\
\supit{h}Saint Mary's University,
   Halifax, NS B3H 3C3,
   Canada; \\
\supit{i}Universit\'e de Montr\'eal,  Montr\'eal, Qu\'ebec H3T 1J4, Canada \\
}

\authorinfo{Further author information: (Send correspondence to M.D.A.)\\M.D.A.: E-mail: mda@roe.ac.uk, Website: http://www.roe.ac.uk/atc/projects/scubatwo/}

\begin{document} 
  \maketitle 

{\bf Copyright 2004 Society of Photo-Optical Instrumentation Engineers.\\}
This paper will be published in SPIE conference proceedings volume 5498, 
``Millimeter and Submillimeter Detectors for Astronomy II.''  and is made available as 
an electronic preprint with permission of SPIE.  One print or electronic 
copy may be made for personal use only.  Systematic or multiple reproduction, 
distribution to multiple locations via electronic or other means, duplication 
of any material in this paper for a fee or for commercial purposes, or 
modification of the content of the paper are prohibited.

\begin{abstract}
SCUBA-2, which replaces SCUBA (the Submillimeter Common User Bolometer
Array) on the James Clerk Maxwell Telescope (JCMT) in 2006, is a 
large-format bolometer array for submillimeter astronomy. Unlike previous
detectors which have used discrete bolometers, SCUBA-2 has two
dc-coupled, monolithic, filled arrays with a total of $\sim$10,000
bolometers. It will offer simultaneous imaging of a 50 sq-arcmin field
of view at wavelengths of 850 and $450\rm\ \mu m$. SCUBA-2 is expected to
have a huge impact on the study of galaxy formation and evolution in the
early Universe as well as star and planet formation in our own Galaxy.
Mapping the sky to the same S/N up to 1000 times faster than SCUBA, it
will also act as a pathfinder for the new submillimeter interferometers
such as ALMA. SCUBA-2's absorber-coupled pixels use superconducting
transition edge sensors operating at 120~mK for 
performance limited by the sky background photon noise. The
monolithic silicon detector arrays are deep-etched by the Bosch process
to isolate the pixels on silicon nitride membranes. Electrical
connections are made through indium bump bonds to a SQUID time-domain multiplexer (MUX). We give an overview
of the SCUBA-2 system and an update on its status, and describe some of
the technological innovations that make this unique instrument possible.
\end{abstract}



\newcommand {\Section}[1]{Section~\ref{#1}}

\newcommand {\Figure}[1]{Figure~\ref{#1}}

\newcommand {\Table}[1]{Table~\ref{#1}} 

\newcommand {\page}[1]{page~\pageref{#1}}

\newcommand {\Equation}[1]{Equation~\ref{#1}}

\newcommand {\arcdeg}{^\circ}
\def\plotfiddle#1#2#3#4#5#6#7{\centering \leavevmode
\vbox to#2{\rule{0pt}{#2}}
    \includegraphics{#1}}


\def\captionfigure#1[#2]#3{
 \def\captionlabel{#1}
 \def\captionlistentry{#2}
 \def\captionheading{#3}
 \begin{figure}}

\def\endcaptionfigure{
 \spacing{1}
 \caption [\captionlistentry]{\captionheading}
 \label {\captionlabel}
 \end{figure}}

\def\sqr#1#2{{\vcenter{\vbox{\hrule height.#2pt
        \hbox{\vrule width.#2pt height#1pt \kern#1pt
          \vrule width.#2pt}
        \hrule height.#2pt}}}\relax}
\def\square{\mathchoice\sqr45\sqr45\sqr{3.1}4\sqr{2.5}4}

\section{INTRODUCTION}
SCUBA (the Submillimetre Common User Bolometer Array) on the James Clerk Maxwell Telescope has been one of the most successful instruments ever built for a ground-based telescope\cite{Holl99}.  Examples of scientific highlights from SCUBA are the discovery of an unknown population of dusty star-forming galaxies\cite{Smai97} and detailed observations of dusty debris disks around nearby stars where planets are forming\cite{Holl03}.  SCUBA-2 will leverage new technologies for a revolutionary improvement in sensitivity and mapping speed.  The science case for SCUBA-2 is discussed in detail elsewhere\cite{SPIE02WSH}.

Less than 1~square degree of sky has been mapped in the sub-millimeter to any great depth (e.g. close the the extragalactic confusion limit). SCUBA-2 will be primarily a survey instrument that will
map large areas of sky.  The new sub-millimeter interferometers like ALMA and the SMA are limited as survey instruments by their narrow fields of view.  SCUBA-2 will act as a pathfinder for these instruments, filling the JCMT's $8\times8$-arcmin field of view with dc-coupled pixels.  
SCUBA-2's field of view is 12 times that of SCUBA.  Also, SCUBA is undersampled by a factor of four, taking 64 pointings (a total of 128 s) for SCUBA to produce a fully-sampled image at 850 and $450\rm \mu m$ simultaneously.  SCUBA-2 will produce a fully-sampled image in just one pointing at $850\rm\ \mu m$ or four pointings at $450\rm\ \mu m$.  This means that SCUBA-2 will be able to map large areas of sky about 1000 times faster than SCUBA (see Table~\ref{table:1}).  

With fully-sampled image planes and no sky chopping, SCUBA-2 will have better image fidelity and map dynamic range than SCUBA.  Imaging at two colours simultaneously, SCUBA-2 will allow us
to study the physical properties of dust emission and to exploit the better angular resolution at $450\rm\ \mu m$ when weather permits.      
SCUBA-2 will also carry out deep imaging to the confusion limit in only a few hours.

\begin{table*}[bht]
\caption{Estimated performance of SCUBA-2.}
\label{table:1}
\renewcommand{\tabcolsep}{1pc} 
\renewcommand{\arraystretch}{1.2} 
\newcommand{\m}{\hphantom{$-$}}
\begin{tabular}{@{}lcccc}
\hline
Parameter           &\multispan2{\hfill$450\rm\ \mu m$\hfill} &\multispan2{\hfill$850\rm\ \mu m$\hfill}\\
&SCUBA&{\bf SCUBA-2}&SCUBA&{\bf SCUBA-2}\\
\hline
Per-pixel NEFD ($\rm mJy/\sqrt{Hz}$)   & 400& {\bf 80} & 90& {\bf 25} \\
Point source NEFD ($\rm mJy/\sqrt{Hz}$)  & 400& {\bf 113} & 90& {\bf 21} \\
Point source extraction speed relative to SCUBA   & 1& {\bf 13} & 1& {\bf 18} \\
Large area mapping speed relative to SCUBA   & 1& {\bf 850} & 1& {\bf 1250} \\
\hline
\end{tabular}\\[2pt]
\end{table*}

\section{DETECTORS}
\subsection{Pixel Design}
\begin{figure}[htb]
\plotfiddle{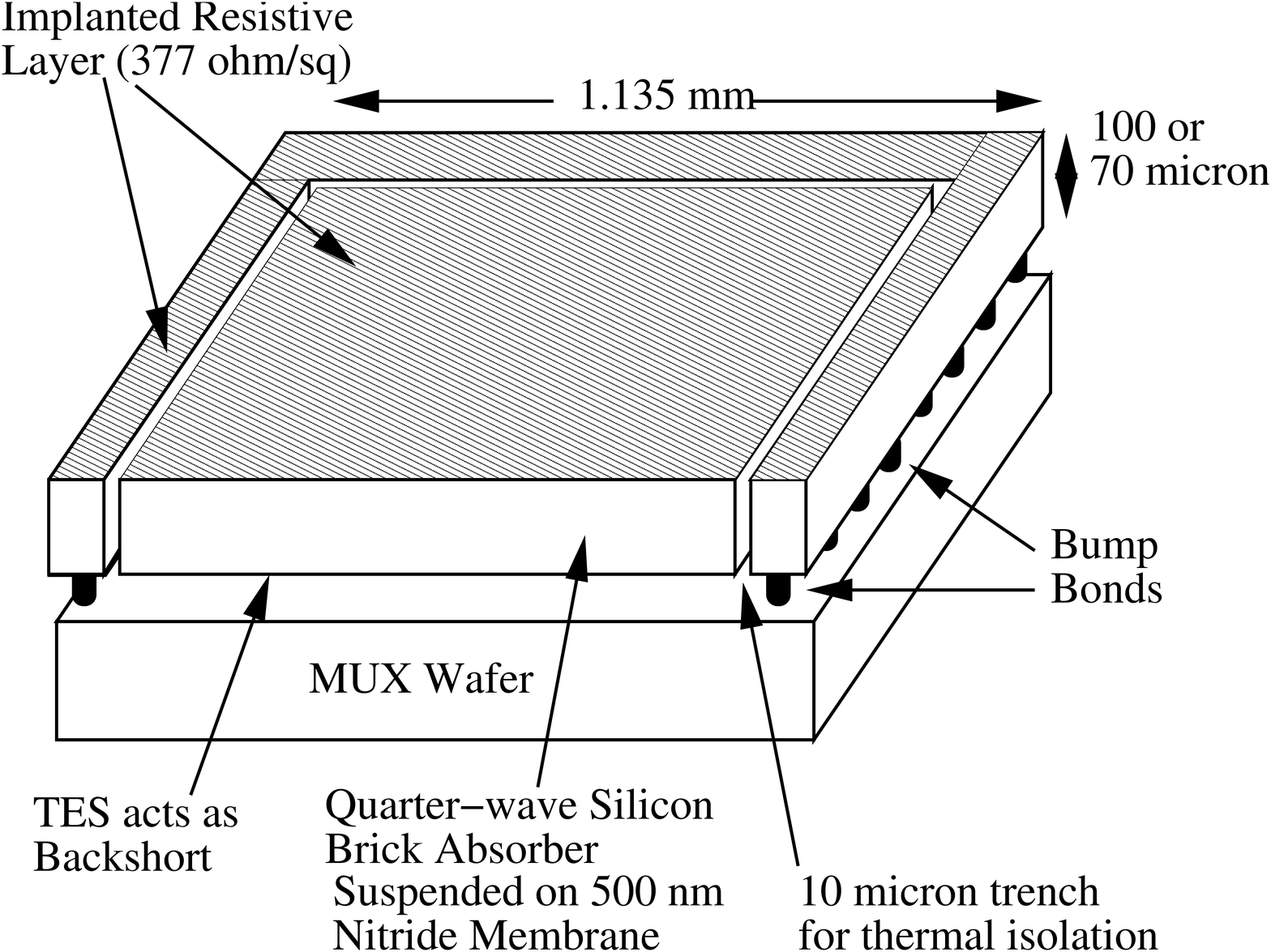}{273.957pt}{0}{25.3}{25.3}{-183.060pt}{-0.000pt}
\caption{SCUBA-2 unit cell.}
\label{fig:pixel}
\end{figure}

The layout of a SCUBA-2 pixel is shown in Figure~\ref{fig:pixel}.  The design incorporates several novel features.  The use of superconducting transition edge sensors (TES) and close integration between the absorber and thermometer result in sensitive, low-noise detectors that are easy to make into arrays, while the use of pixel heaters gives the detectors the power handling required to cope with the variable sub-millimetre sky background.  

\subsection{Transition Edge Sensors}
The temperature rise due to incident radiation is detected by a TES.  The TES is voltage biased so that it operates in the regime of strong negative electrothermal feedback\cite{KentsThesis}.  If the temperature drops, so does the resistance of the TES.  Since it is biased at constant voltage, this means that the current, and hence the Joule power, will increase, heating up the TES.  Conversely, if the temperature increases the resistance will increase, reducing the current, and thus the Joule heating.  This means that the TES operates at a bias point that is in a stable equilibrium.  Thus, the TES is self-biasing.  There is no need for a temperature controller to ensure that it remains at the correct bias point.  Also, the electrothermal feedback cancels out temperature fluctuations which has the effect of suppressing the Johnson noise.  Because a TES is a low-impedance device it is not very susceptible to microphonics.  SCUBA's thermistors, in contrast, were affected by vibrations from the JCMT's secondary mirror unit (SMU).  This meant that flat-fielding by dithering the SMU was not feasible for SCUBA.  We can use such a mode for SCUBA-2 (see \Section{sect:DREAM}) because its low-impedance detectors are insensitive to microphonics.  Another advantage of using a TES as a thermometer is that it is much more sensitive than the thermistors used in SCUBA.  The TES films in SCUBA-2 are Mo/Cu proximity-effect bilayers.  The transitions of the bilayers can be made as sharp as 1--2~mK for high sensitivity.  We can also tune the transition temperature ($T_c$) of the films to the desired value by choosing the deposition properties.  

\subsection{Absorber-Coupled Pixels}
\begin{figure}[htb]
\vspace{9pt}
{\centering \leavevmode \vbox to125.169pt{\rule{0pt}{125.169pt}
\includegraphics{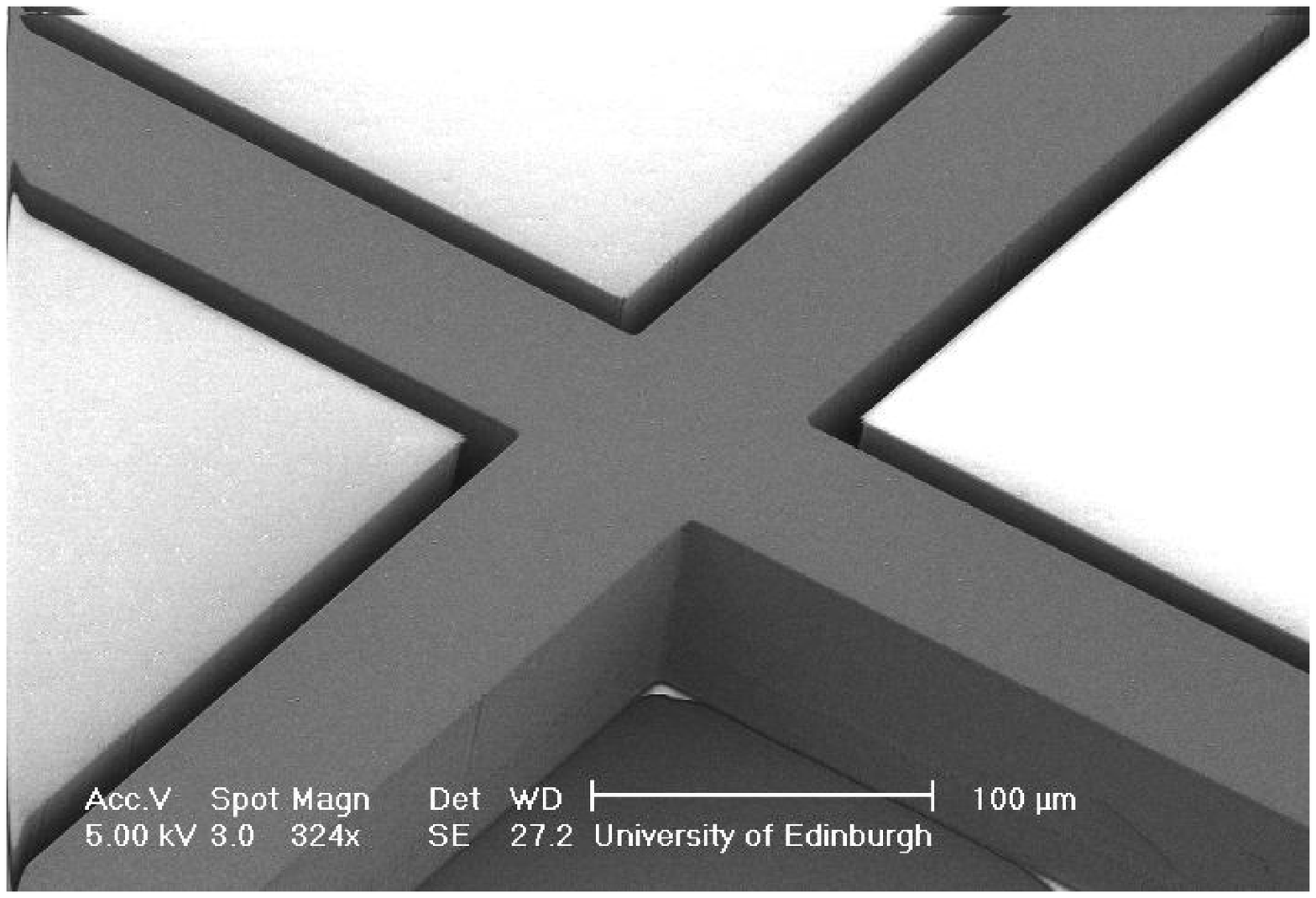}
\includegraphics{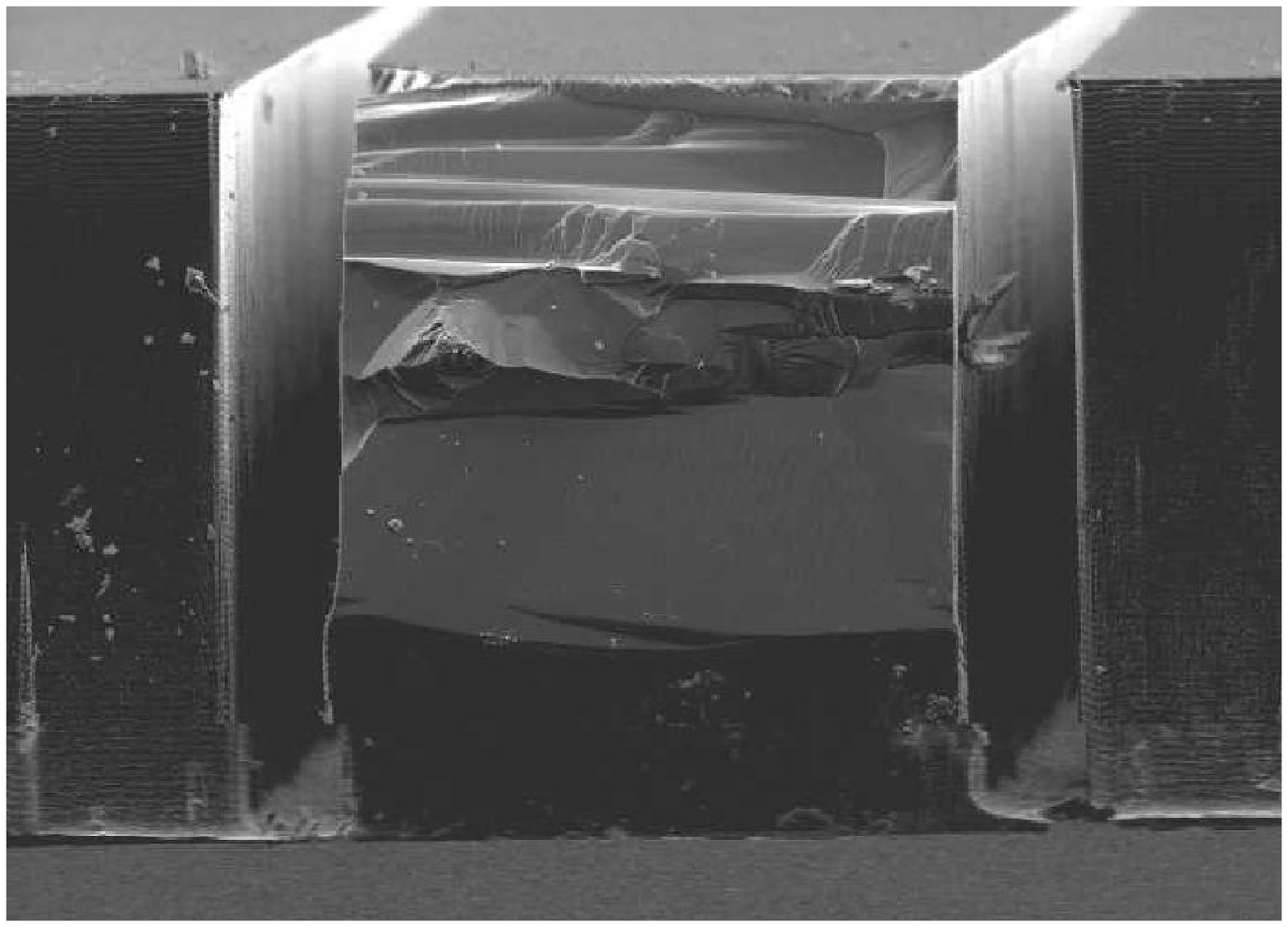}
}}
\caption{Left: SEM image showing deep-etched silicon quarter-wave bricks and walls.  The lighter colour of the bricks is caused by charge buildup.  The bricks are suspended on silicon nitride membranes and are thus less well grounded than the walls.  Right: SEM image showing a $10\rm\ \mu m$ trench.  
The etch stopped 
on the nitride membrane.  The trench is $100\rm\ \mu m$ deep.}
\label{fig:saltire}
\end{figure}

Another feature of the SCUBA-2 detectors is the close coupling between the absorbers and the TES.  The TES forms the backshort of a quarter-wave silicon ``brick'' with an implanted absorber at the front.  This behaves essentially as an antireflection coating.  The surface of the brick is ion-implanted to $377\ \Omega/\square$ to match free space and to dissipate incident radiation as heat.  The detectors are formed from two 3'' wafers diffusion-bonded together.  The upper ``handle'' wafer provides mechanical support during the wafer thinning and hybridization steps while the lower (detector) wafer is used to form the quarter-wave silicon bricks.  The detector wafer is thinned down to $63\rm\ \mu m$ or $100\rm\ \mu m$, depending on the detection wavelength, and a silicon nitride membrane is deposited on the back.  This silicon nitride membrane provides the weak thermal link from the bolometer to the heat bath.  The isolating trenches around the absorber are $10\rm\ \mu m$ wide and are made by the Bosch process\cite{Park04} (see \Figure{fig:saltire}).  Underneath each brick is a TES thermometer which also acts as a backshort.  Finite-element electromagnetic modeling predicts that the SCUBA-2 absorbers will have efficiencies of 88\%\ at $850\rm\ \mu m$ and 93\%\ at $450\rm\ \mu m$\cite{SPIE04Poster}.

The pixels in the prototype are different from the original design\cite{SPIE02WDD} in which the inter-pixel walls extended $381\rm\ \mu m$ above the quarter-wave bricks, forming a mechanical support grid.  We had originally intended to complete the micromachining of the detector wafer and then hybridize it to the MUX wafer.  However, this turned out to be impractical.  The detector wafer would still be too fragile for hybridization, even with the support grid.  We decided instead to do the deep-etch micromachining after hybridization.  This meant that the mechanical support grid was now redundant.  Also, electromagnetic modeling showed that the grid degraded the detection efficiency.  Thus, we now remove the handle wafer completely after hybridization.

\subsection{Pixel Heaters}
Using TES bolometers for sub-millimetre astronomy presents a challenge with dynamic range.  The sources we observe are small perturbations on a bright, fluctuating background due to emission from water vapour in the atmosphere.  If the sky power becomes too large there is a danger that the TES will be driven normal.  In order to ensure that we can operate the SCUBA-2 detectors over the range of expected sky powers at the JCMT we have included a heater on each pixel.  These heaters are wired in series and the heating power is controlled on a slow servo to the average power over the whole sub-array.  This increases the power handling of the detectors because it allows them to operate at about the same bias power no matter what the sky background conditions are.  This has the advantage that we can choose to operate the TES higher up in the transition where there is less excess noise\cite{Linde04}, but also less overhead for power handling.

\subsection{Pixel Performance}
Single-pixel test devices were used to verify the pixel design.  A summary of results is shown in \Table{table:performance}.  The transition temperatures ($T_c$) of the devices were close to the required values.  We determined the optimum bias conditions and verified the heater performance.  The noise contribution from the heaters was much less than the expected sky noise.  Both the 850 and $450\rm\ \mu m$ pixels achieved the required noise performance and power handling.  The $450\rm\ \mu m$ devices were a little faster than we would like.  The main constraint on the speed of the TES is that the effective time constant must be greater than the $L/R$ time constant of the readout circuit to ensure stability against oscillations.  We cannot simply reduce the inductance, and hence the time constant, of the circuit because that would require faster sampling to satisfy the Nyquist criterion.  The speed of the $450\rm \mu m$ detectors will be reduced during the development of the first $450\rm\ \mu m$ prototype sub-array. 

\begin{table*}[bht]
\caption{Measured performance of SCUBA-2 test pixels at likely operating points.  it is assumed that the pixels are run with bias resistances of 25 and $45\rm\ m\Omega$ and heater powers of 40 and 200~pW for the 850 and $450\rm\ \mu m$ detectors, respectively.  The absorption efficiencies are derived from electromagnetic modeling\cite{SPIE04Poster}.  The value of $T_c$ quoted for the $850\rm\ \mu m$ detectors is for the first prototype sub-array.}
\label{table:performance}
\renewcommand{\tabcolsep}{1pc} 
\renewcommand{\arraystretch}{1.2} 
\newcommand{\m}{\hphantom{$-$}}
\begin{tabular}{@{}lcccc}
\hline
Wavelength          &\multispan2{\hfill$850\rm\ \mu m$\hfill} &\multispan2{\hfill$450\rm\ \mu m$\hfill}\\
&Goal&{Measured}&Goal&{Measured}\\
\hline
$T_c$ (mK)&$130\pm10$&137&$190\pm10$&193.2\\
Response time (ms)           & 1--2& 0.85 & 1--2& 0.09 \\
Noise-equivalent power ($\rm W/\sqrt{Hz}$) &$<3\times10^{-17}$&$3.5\times10^{-17}$ &$<1.5\times10^{-16}$&$1.1\times10^{-16}$ \\
Absorption Efficiency (\%)           &$\ge75$ & 88 &$\ge75$ & 93 \\
\hline
\end{tabular}\\[2pt]
\end{table*}

\section{SUB-ARRAYS}
\subsection{Sub-array Design}
\begin{figure}[htb]
\vspace{9pt}
\plotfiddle{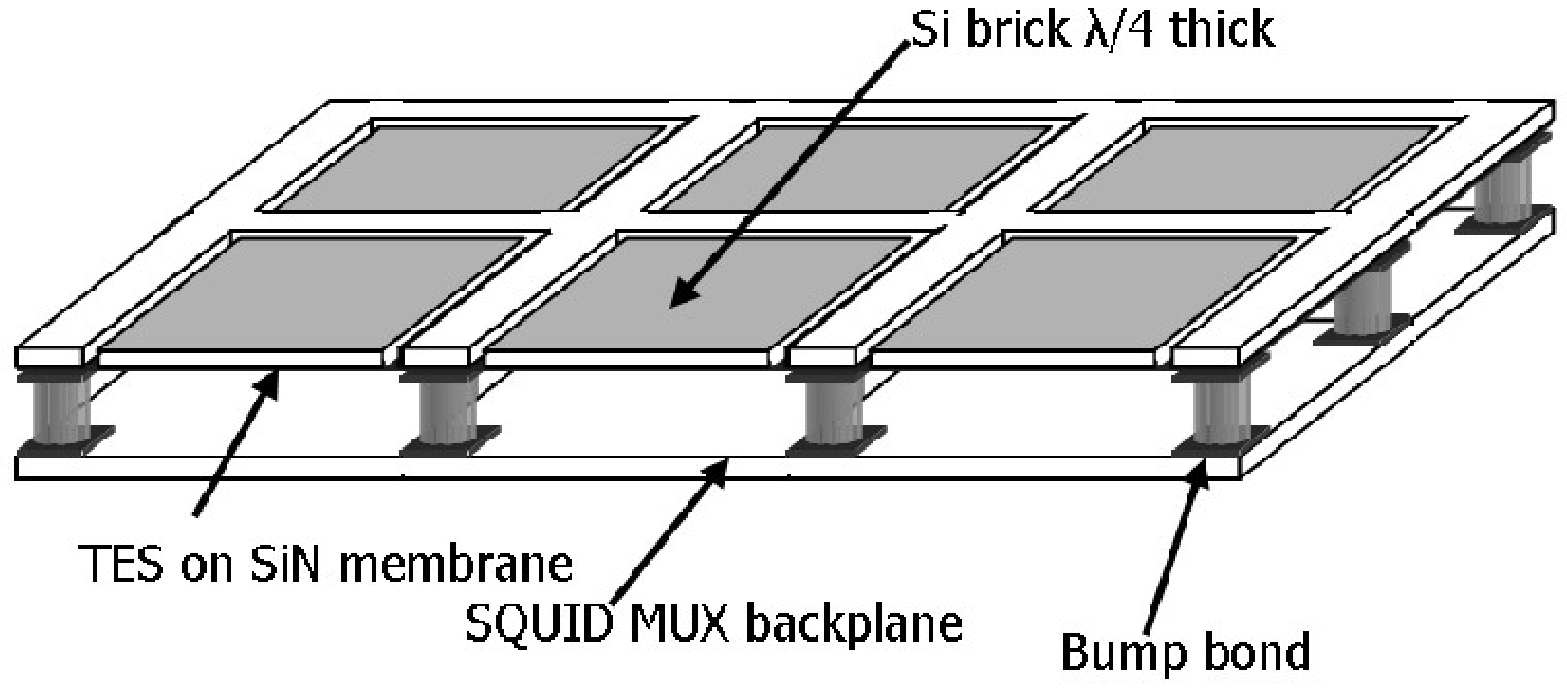}{170.543pt}{0}{78.2}{78.2}{-239.386pt}{-224.522pt}
\caption{Representation of a group of detector pixels within a $40\times32$ sub-array.  The array pitch is 1.135~mm.  The interpixel walls are $50\rm\ \mu m$ thick and the same height as the quarter-wave bricks..}
\label{fig:array}
\end{figure}

\begin{figure}[htb]
\plotfiddle{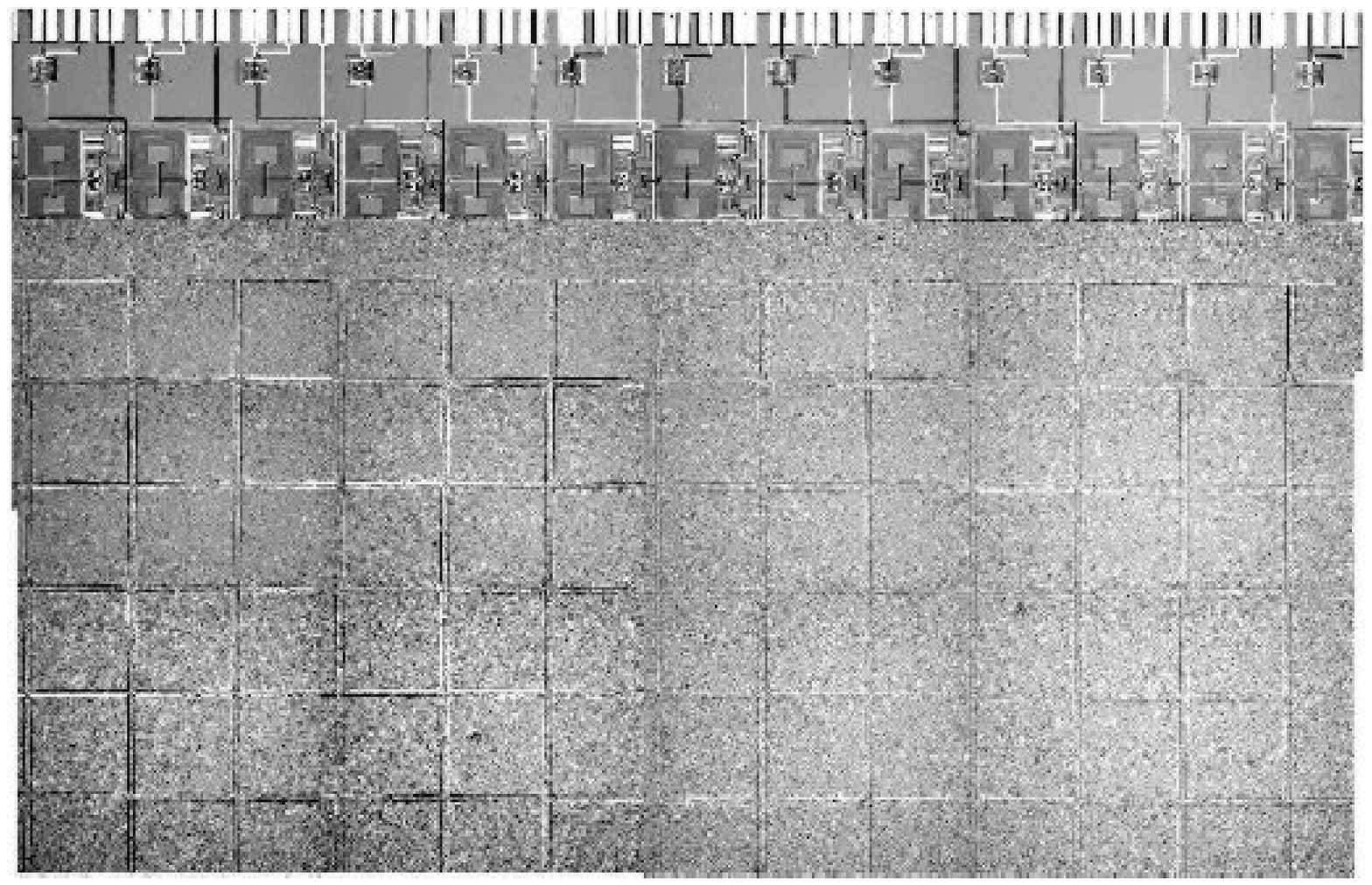}{234.692pt}{0}{78.2}{78.2}{-239.386pt}{-192.448pt}
\caption{Optical image of the $850\rm\ \mu m$ prototype sub-array.  This is a mosaic of photographs taken at $25\times$ magnification.  At the top of the picture the wire bond pads are visible.  Below them are the second stage SQUIDs (one for each column).  The 1~mm square trenches that isolate the pixels on the detector wafer are clearly visible.  The larger-scale shading pattern on the image is an artifact of the mosaicing.}
\label{fig:prototype}
\end{figure}

The focal plane is populated by four sub-arrays of $40\times32$ pixels, each of which is fabricated from 3'' Si wafers.  The layout of a sub-array is shown in Figure~\ref{fig:array}.  The two new technologies that make the SCUBA-2 sub-arrays feasible are indium bump-bond hybridization and the SQUID MUX\cite{LTD10KDI}.  The sub-arrays have been shrunk to $40\times32$ pixels from the original $40\times40$\cite{LTD-9} for more reliable patterning of the MUX on a 3'' wafer.  The first $850\rm\ \mu m$ prototype sub-array has been sucessfully fabricated (see \Figure{fig:prototype}).  At the time of writing it is being integrated into a sub-array module for testing.  The sub-array module is described in the next section.

\subsection{Bump-Bond Hybridization}
\begin{figure}[htb]
\plotfiddle{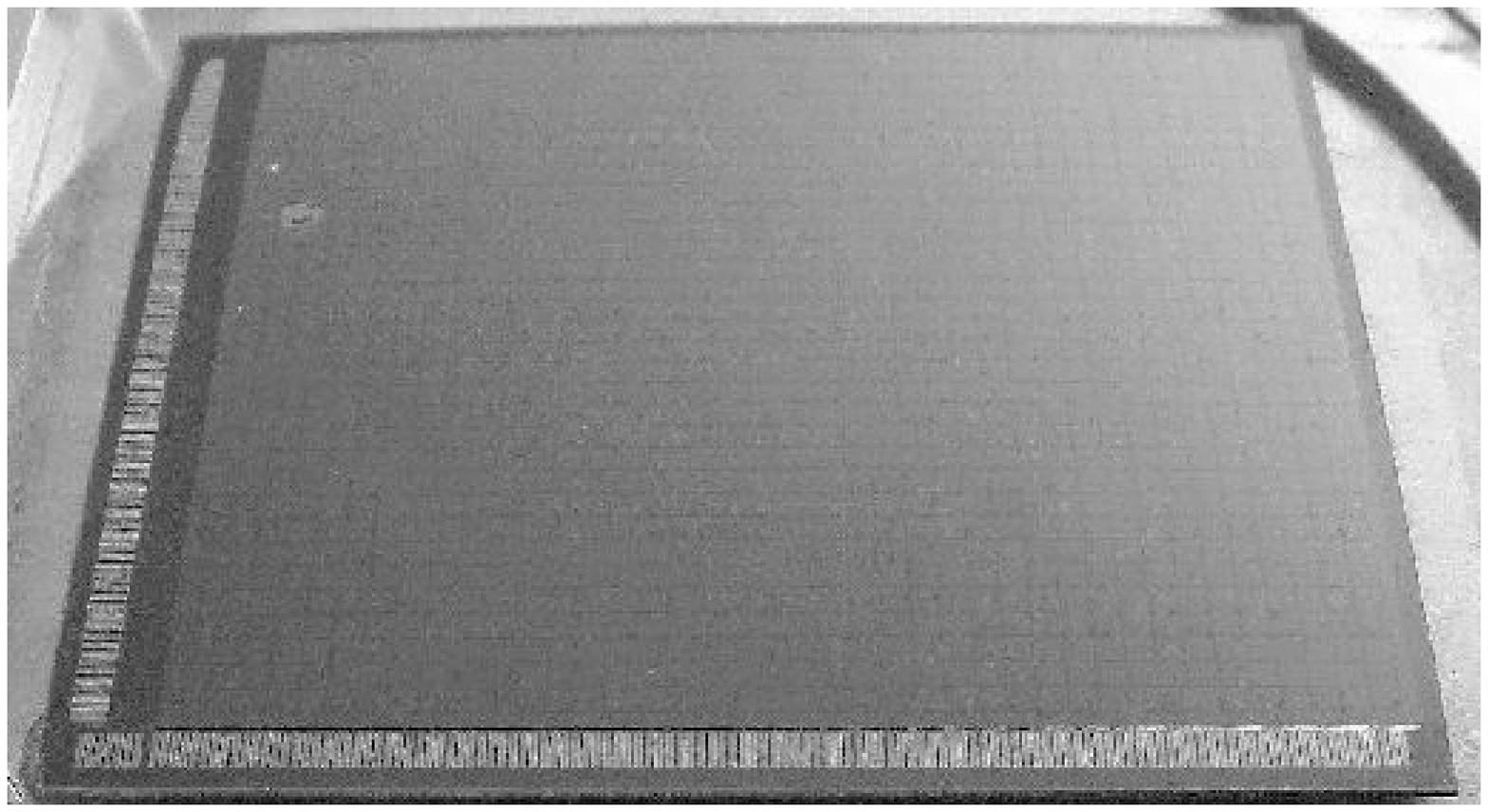}{198.706pt}{0}{78.2}{78.2}{-239.386pt}{-210.441pt}
\caption{Hybridized dummy similar to the prototype sub-array.  The size of the subarray is $41\times52\rm\ mm$.}
\label{fig:dummy}
\end{figure}

The electrical and thermal connections to the MUX are made through indium bump bonds.  Indium has the advantage that the bumps can be formed by cold welding so that we avoid elevated temperatures that might damage the SQUID circuitry.  However, indium has two disadvantages.  It is superconducting at $100\rm\ mK$, and it is not very strong.
In order to ensure that the heat is transported out of the pixels there are 79 bumps per pixel.  There is also a ``forest'' of bump bonds around the edges of the sub-array to provide extra mechanical strength.  The bump bonds also provide electrical connections to the MUX wafer.
\Figure{fig:dummy} shows a hybridized dummy sub-array that was made while verifying the process.

\subsection{SQUID Multiplexer}
\begin{figure}[htb]
\plotfiddle{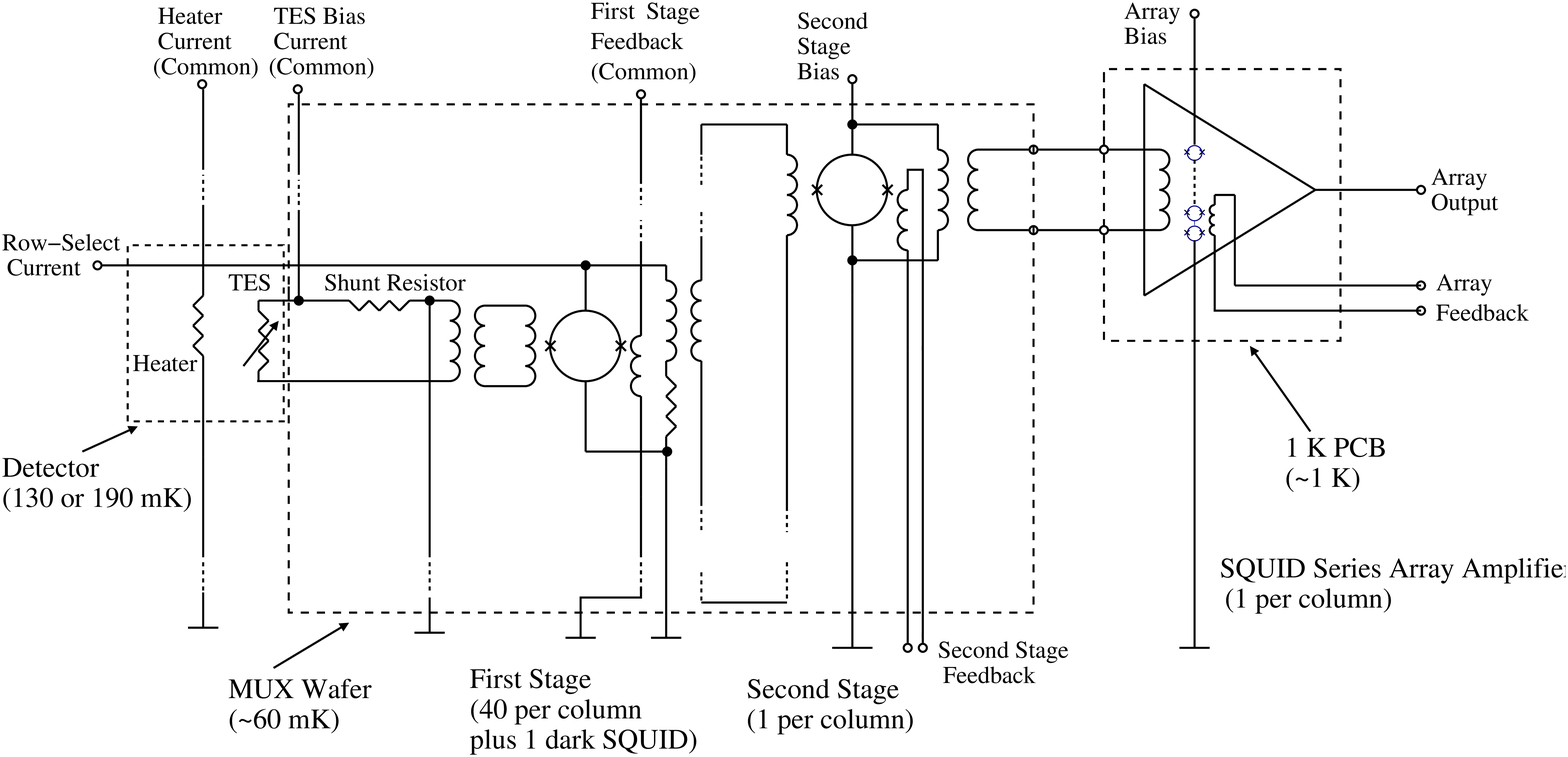}{176.927pt}{0}{23.6}{23.6}{-183.060pt}{-0.000pt}
\caption{Simplified diagram of the SCUBA-2 readout.  The dummy SQUID that cancels out crosstalk in the first stage is not shown.  Note that all grounding is done outside the cryostat.  For every line that enters the instrument there is a return line.}
\label{fig:mux}
\end{figure}

The indium bump bonds connect each TES to a corresponding SQUID which detects the change in conduction current due to the incident radiation and amplifies the signal.   There is a balanced pair of SQUIDs under each pixel.  One of the SQUIDs in a pair is unbiased and cancels out crosstalk in unaddressed pixels as the SQUID bias and feedback are switched from row to row.  There is also a $41^{\rm st}$ dark SQUID in each column which is used to cancel out drift ($1/f$ noise) in the downstream electronics.  A second stage of SQUIDs provides further amplification.  There is one second-stage SQUID per column.  A simplified schematic of the SCUBA-2 readout is shown in \Figure{fig:mux}.
 A full $32\times40$ wafer-scale multiplexer has been demonstrated\cite{SPIE04Carl}.  Resonances seen in previous versions of the MUX have been eliminated.  96\%\ of the SQUID critical currents were found to be in range and the critical current uniformity will be improved by monitoring the etch endpoint.  Unfortunately, the dark SQUID does not work on the present devices.  This is due to a layout error which has been corrected for future MUX wafers.  The noise was measured at 4~K as $1.6\rm\ \mu\Phi_0/\sqrt{Hz}$ --- a factor of two within the specification.  Based on these results we proceeeded to manufacture the first prototype sub-array.

\begin{figure}[htb]
\plotfiddle{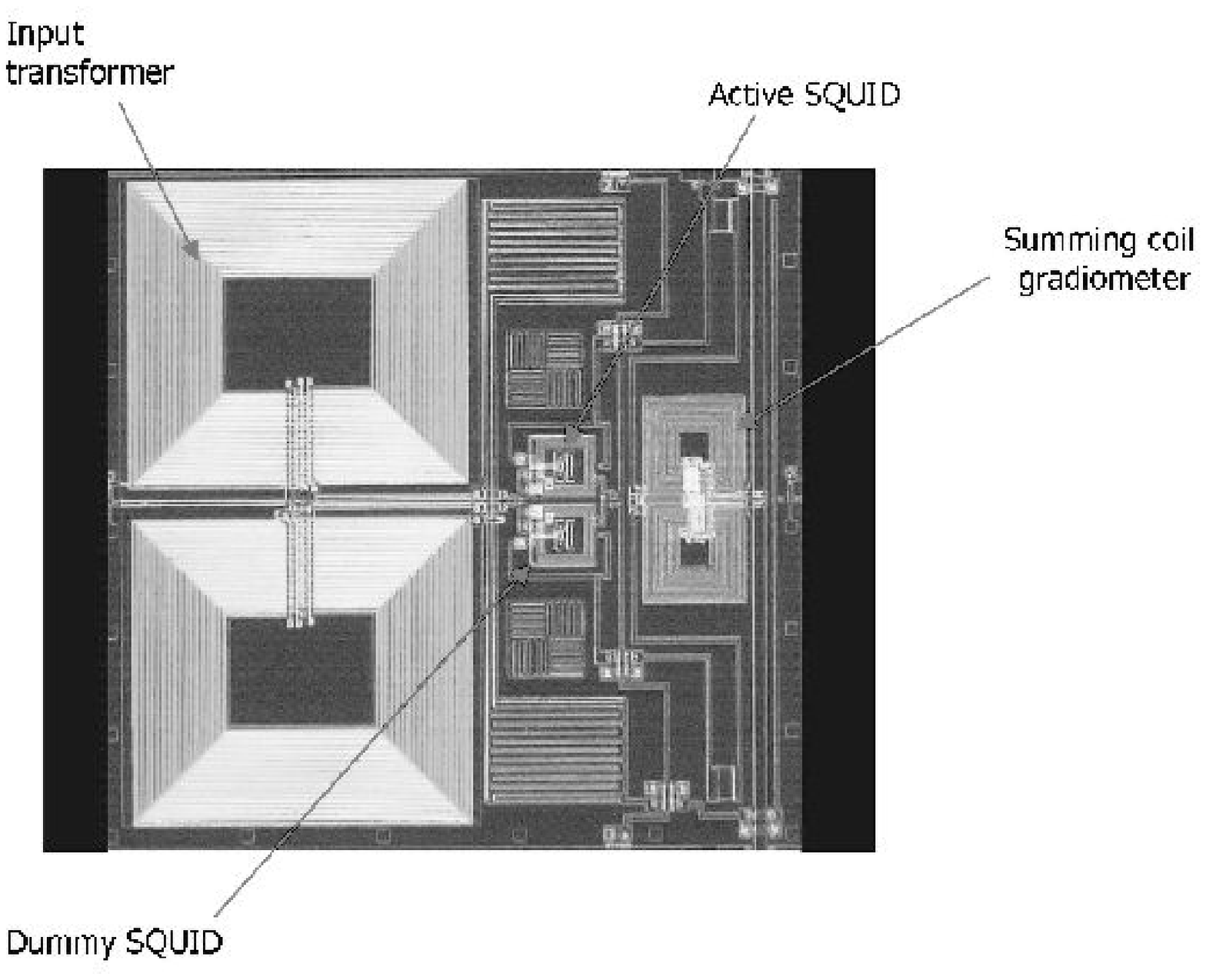}{289.454pt}{0}{78.2}{78.2}{-239.386pt}{-165.067pt}
\caption{SCUBA-2 MUX unit cell.}
\label{fig:muxunitcell}
\end{figure}

\begin{figure}[htb]
\plotfiddle{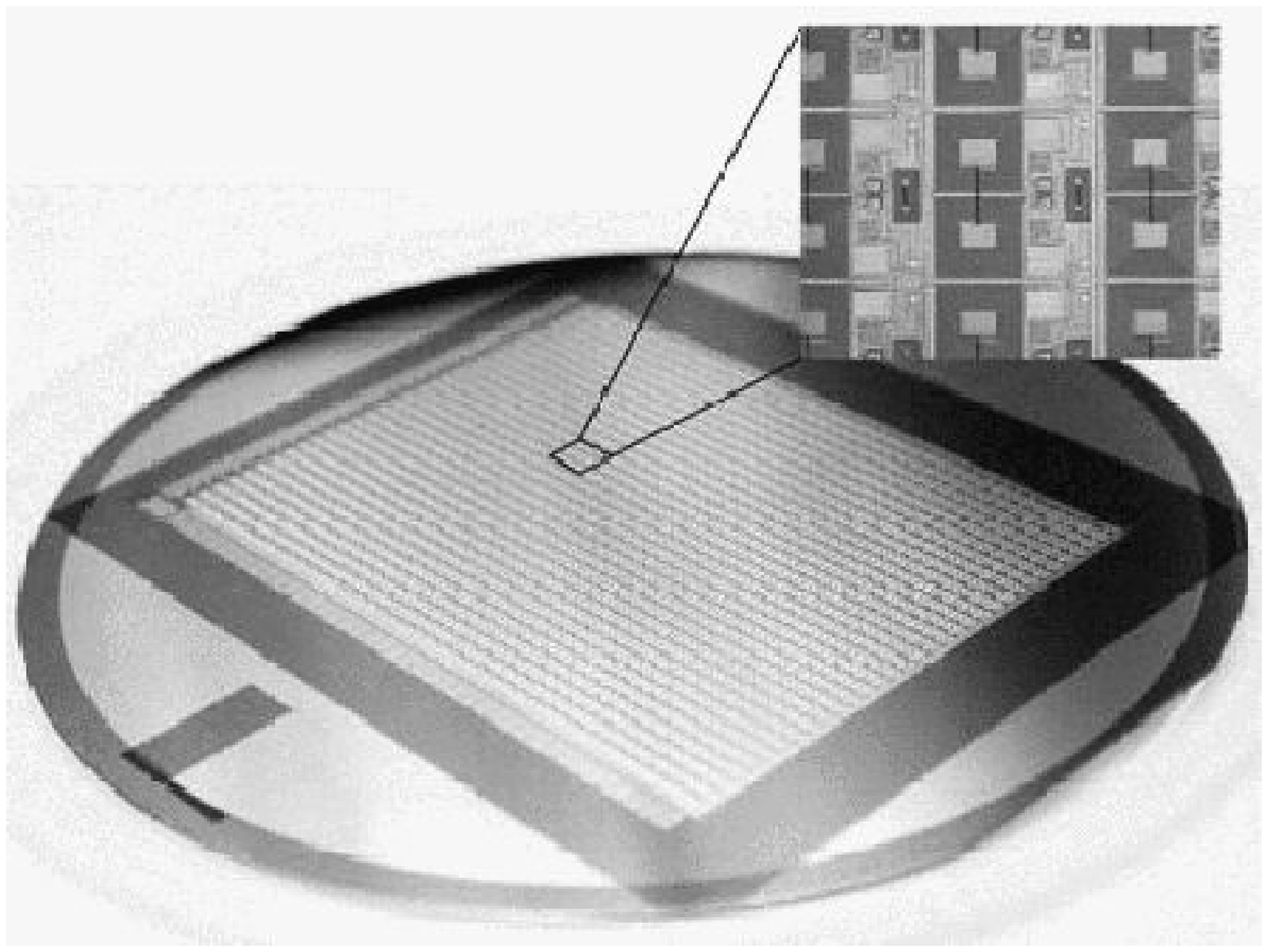}{273.808pt}{0}{78.2}{78.2}{-239.386pt}{-172.890pt}
\caption{The first completed $40\times32$ prototype MUX wafer with detail shown as inset.}
\label{fig:wafermux}
\end{figure}

\section{SUB-ARRAY MODULE}
\subsection{Overview}
\begin{figure}[htb]
\plotfiddle{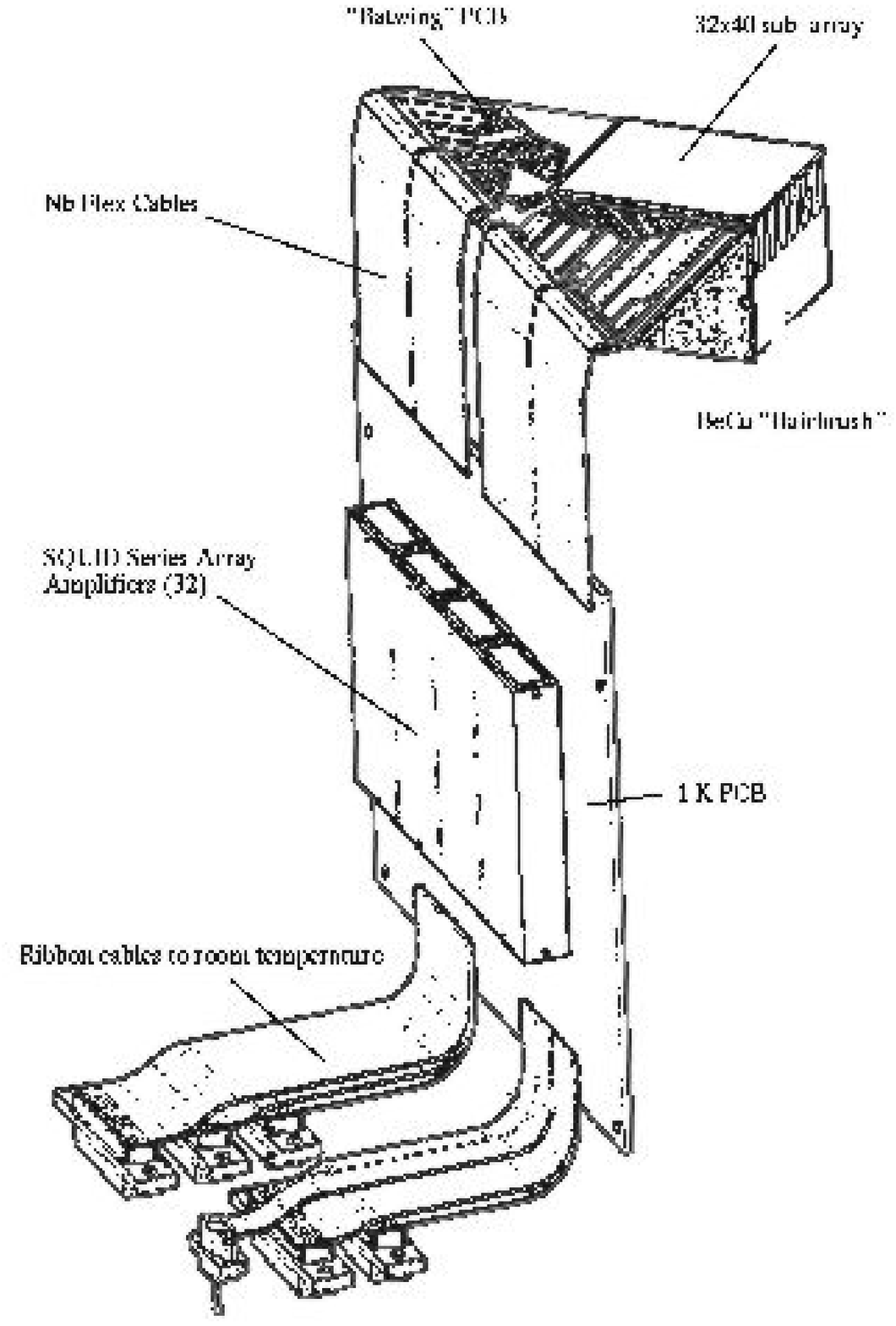}{393.120pt}{0}{61.0}{61.0}{-227.011pt}{-45.172pt}
\caption{SCUBA-2 sub-array module.  The sub-array is connected to the ``batwing'' PCB by aluminium wire bonds.  The Nb flex cables provide a thermal break between the 60~mK and 1~K stages.}
\label{fig:subarraymodule}
\end{figure}

Each sub-array is mounted in a sub-array module (\Figure{fig:subarraymodule}).  The sub-array module provides mechanical support, a cooling path to the dilution refrigerator, and amplification.

\subsection{Chip Holder}
\begin{figure}[htb]
\plotfiddle{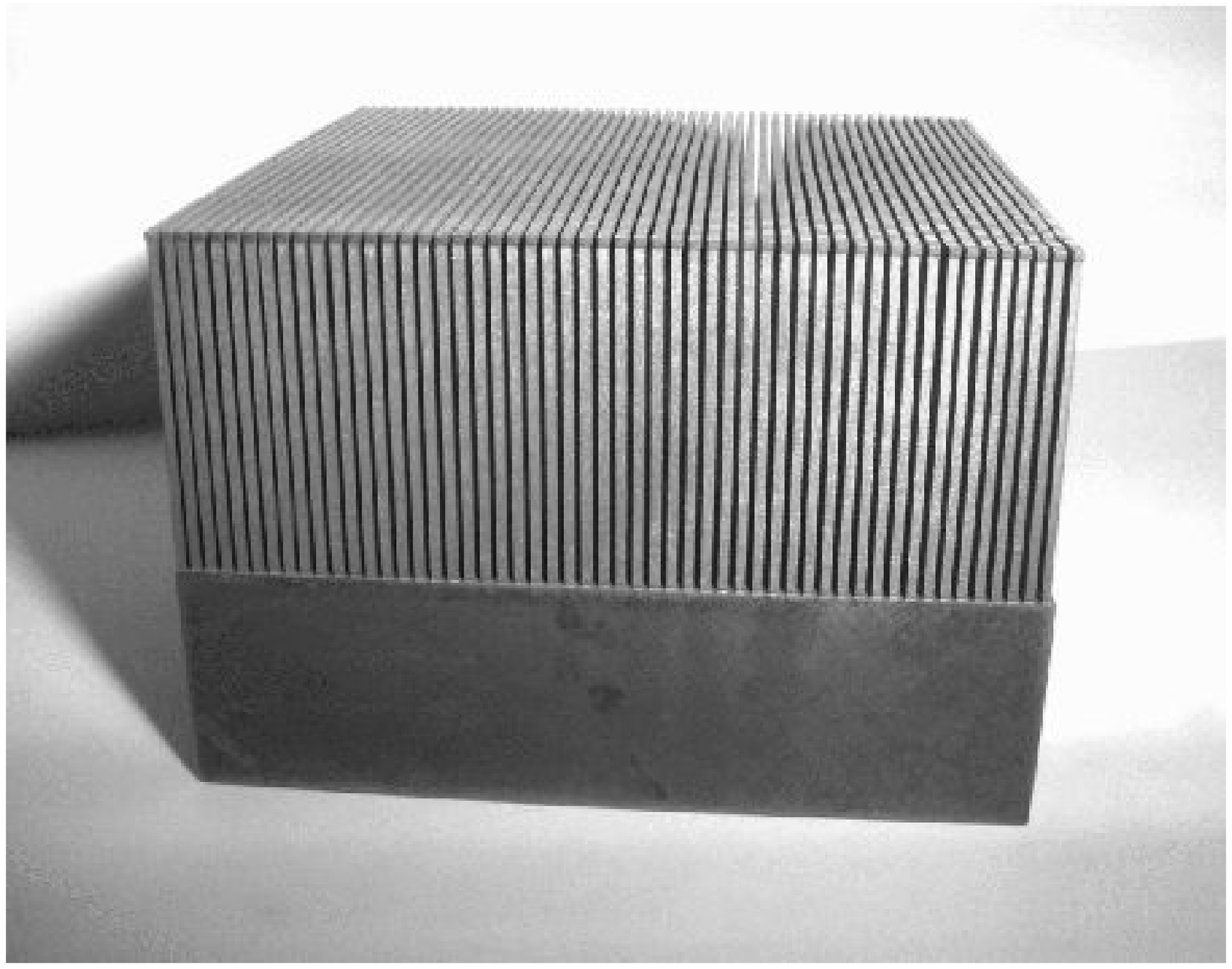}{287.889pt}{0}{78.2}{78.2}{-239.386pt}{-165.849pt}
\caption{Beryllium copper hairbrush 40~mm square.  The tines are 21~mm long with wider pads at the top for better thermal contact.}
\label{fig:hairbrush}
\end{figure}

Cooling such large-format monolithic sub-arrays to $100\rm\ mK$ presents a considerable challenge.  We must make good thermal contact to the back of the MUX wafer, while at the same time relieving stresses caused by differential contraction that could shatter the wafer.  Our solution is to mount the wafer on a ``hairbrush'' structure where contact is made to the wafer by many individual tines, allowing for differential contraction.  There is one tine under each detector pixel and more under the second-stage SQUIDs.  The hairbrush is fabricated by wire erosion from beryllium copper (see \Figure{fig:hairbrush}).  Beryllium copper was chosen for its stiffness.  A copper hairbrush would deform plastically with a chance of breaking after repeated thermal cycles.  We had originally intended to use a gallium-based solder with a low melting point to attach the hairbrush but we found that it was difficult to make it wet other metals.  We are now using Stycast 1266 epoxy to stick each tine of the hairbrush individually to the MUX wafer.  We use a commercial liquid deposition system to deposit a controlled amount of epoxy on each tine.  The hairbrush is bolted on to the copper cold finger that leads to the mixing chamber of the dilution refrigerator.  The thermal interface to the sub-arrays is described elsewhere in these proceedings\cite{Wood04}.

\subsection{Cold Electronics}
\begin{figure}[htb]
\plotfiddle{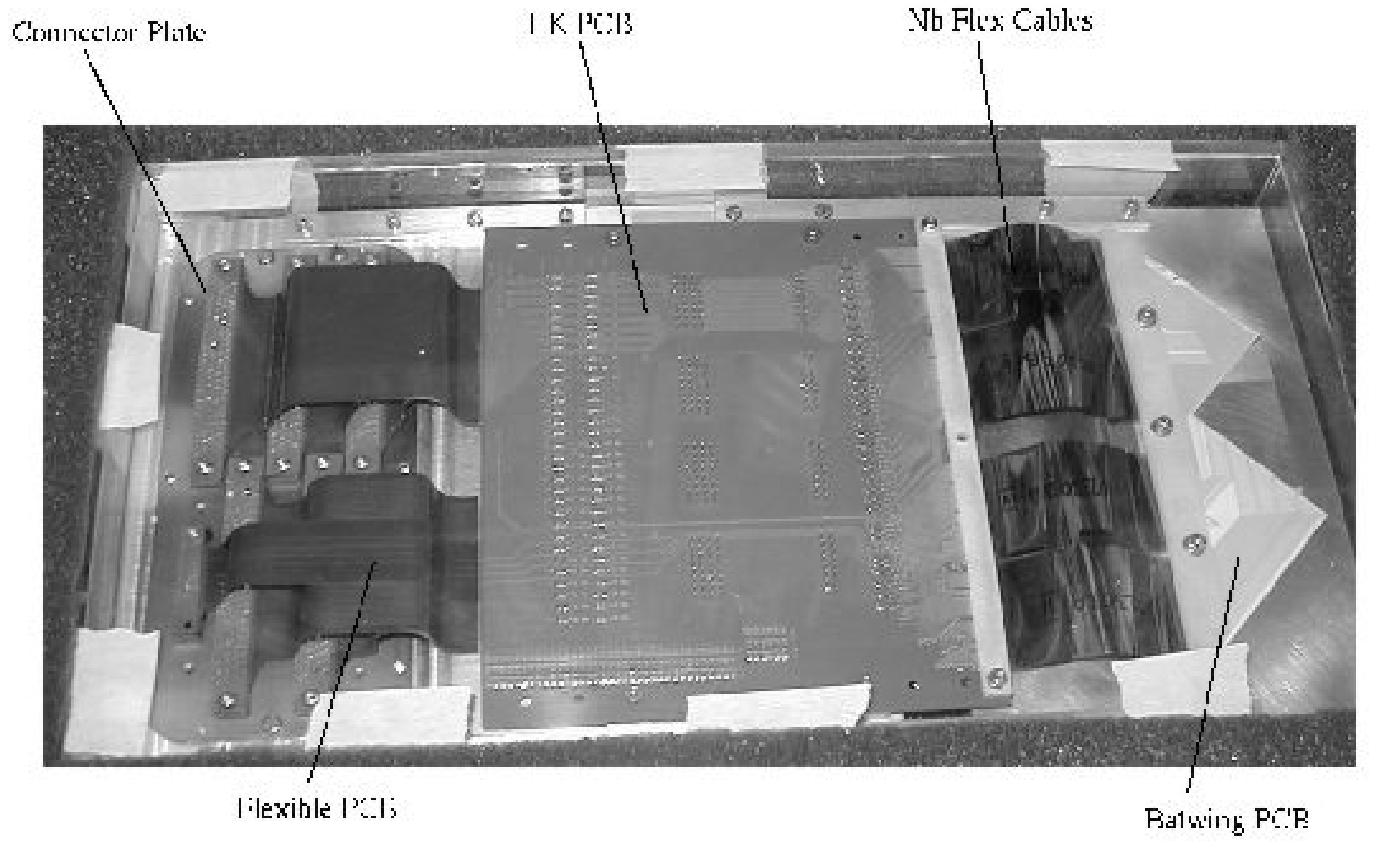}{215.917pt}{0}{78.2}{78.2}{-239.386pt}{-201.835pt}
\caption{Cold electronics assembly for the first prototype sub-array.}
\label{fig:coldelectronics}
\end{figure}

\Figure{fig:coldelectronics} shows an unpopulated sub-array module laid out flat.
The sub-array makes electrical connection to ceramic pcbs (called ``batwing'' pcbs because of their shape) through aluminium wire bonds.  Niobium flex cables carry the signals from the batwing pcb to the 1~K pcb which holds four magnetically-shielded cans, each of which contains eight SQUID series array amplifiers.  Each SQUID series array amplifier amplifies the multiplexed signal from a column.  The flex is attached with an anisotropically-conductive adhesive which conducts electricity vertically but not between traces.  The niobium flexes also provide a thermal break between the 60~mK stage (at the batwing pcbs) and the 1~K stage (the 1~K pcb).  
The 1~K pcb is a multilayer board which has some flexible layers.  These flexible layers go to a connector plate on the 1~K box from which woven ribbon cables connect to the room temperature electronics (see \Section{sect:electronics}).

\section{INSTRUMENT}
\subsection{Focal Plane Unit}
\begin{figure}[htb]
\plotfiddle{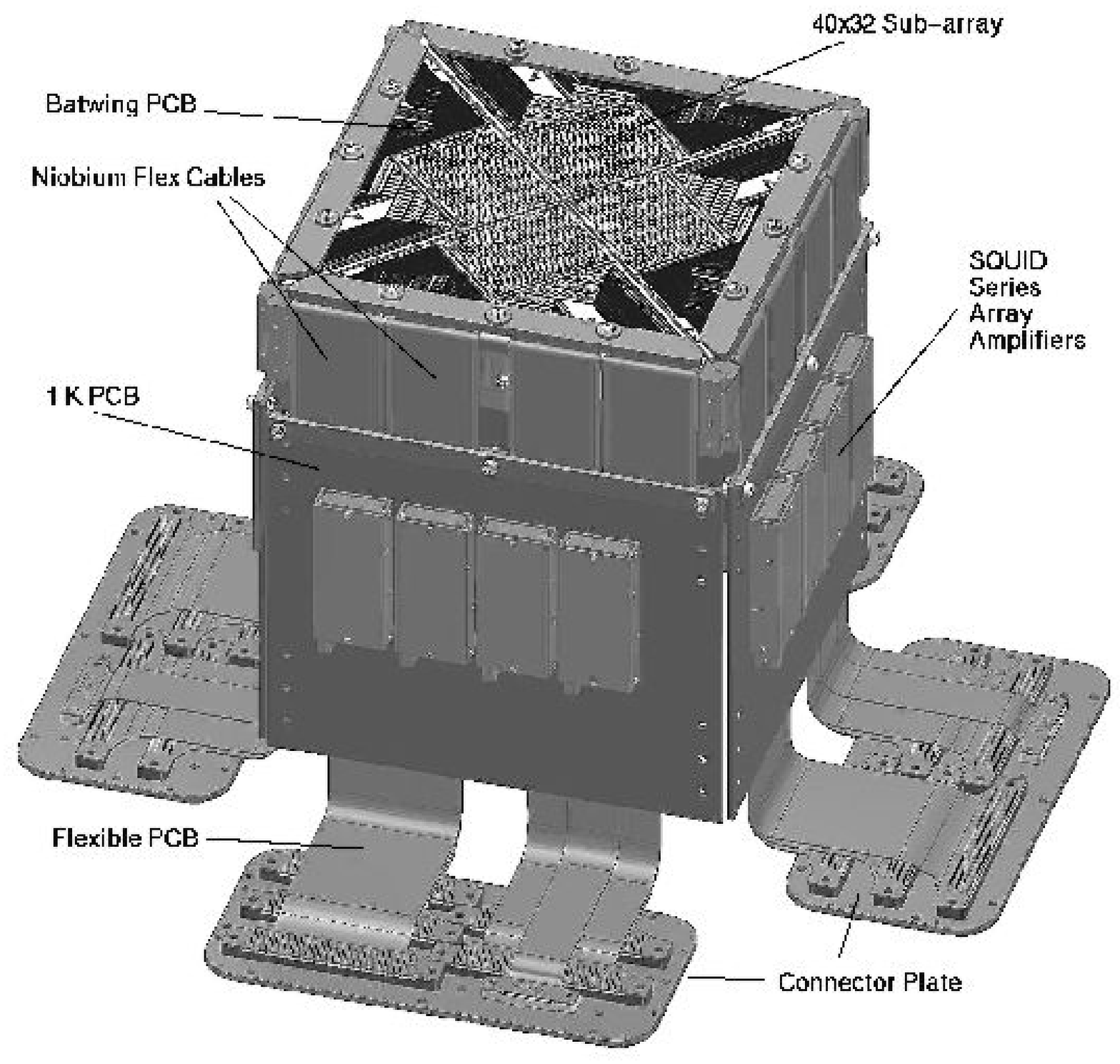}{348.909pt}{0}{78.2}{78.2}{-239.386pt}{-135.339pt}
\caption{SCUBA-2 focal plane unit.}
\label{fig:fpu}
\end{figure}

\begin{figure}[htb]
\plotfiddle{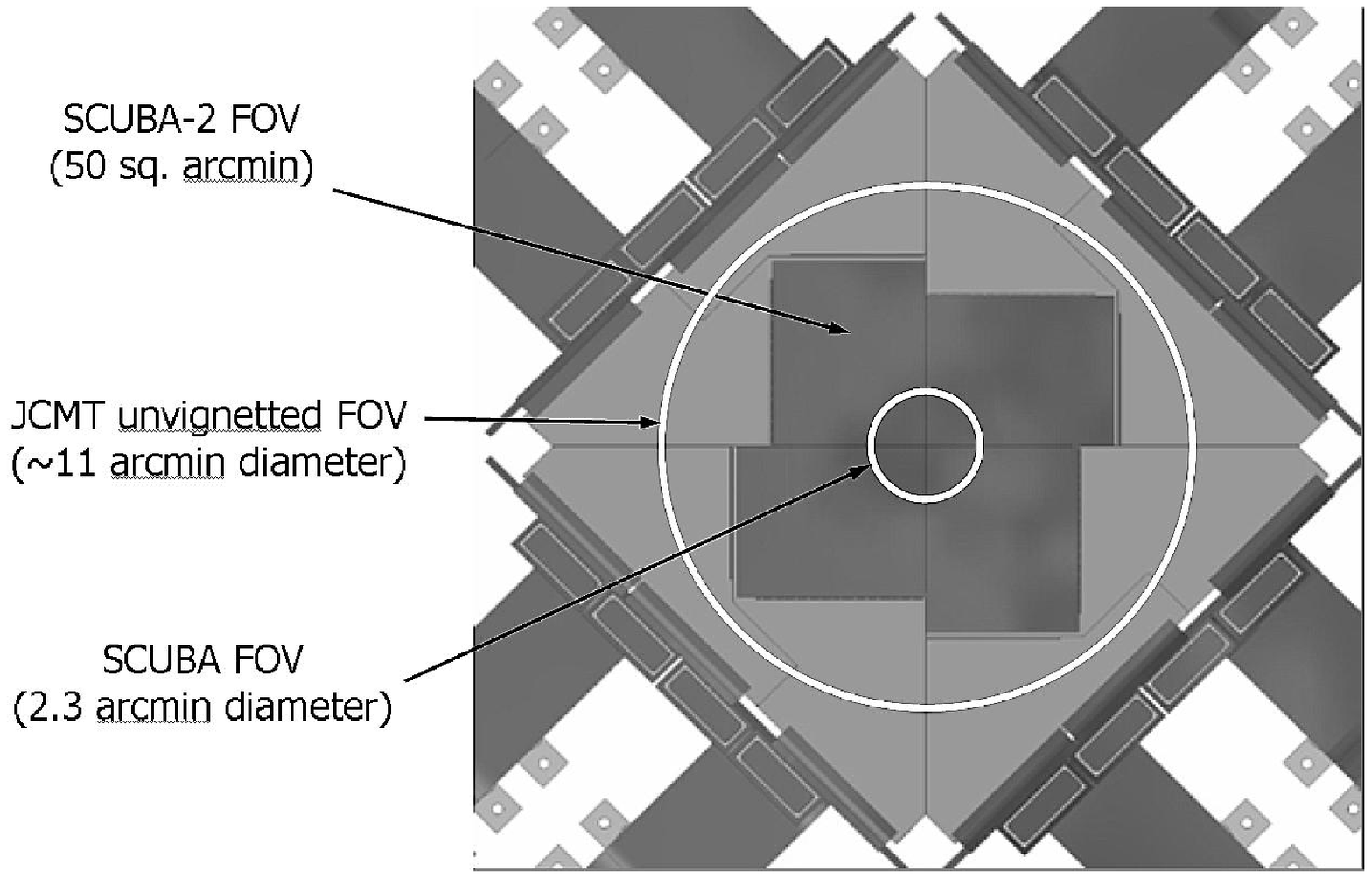}{231.563pt}{0}{78.2}{78.2}{-239.386pt}{-194.012pt}
\caption{Layout of SCUBA-2's focal plane.  The large circle is the field-of-view of the JCMT and the small circle is the field-of-view of SCUBA.}
\label{fig:fputop}
\end{figure}

A high-efficiency dichroic beamsplitter at the 1~K stage splits the beam into into components of wavelength 450 and $850\rm\ \mu m$ and directs them to two focal plane units (FPU).  Each FPU comprises four sub-array modules butted together as shown in Figure~\ref{fig:fpu}.  This arrangement fills most of the 60~arcmin$^2$ field of view of the JCMT with a one-pixel gap between sub-arrays, as shown in \Figure{fig:fputop}.  Since the sub-arrays are now rectangular rather than square, butting four of them together to make a rectangular array would have required us to manufacture two kinds of sub-array (left and right handed), introducing extra cost and complexity.  The arrangement shown in \Figure{fig:fputop} requires only one kind of sub-array.

\subsection{CRYOSTAT}
\subsubsection{Overview}
\begin{figure}[htb]
\plotfiddle{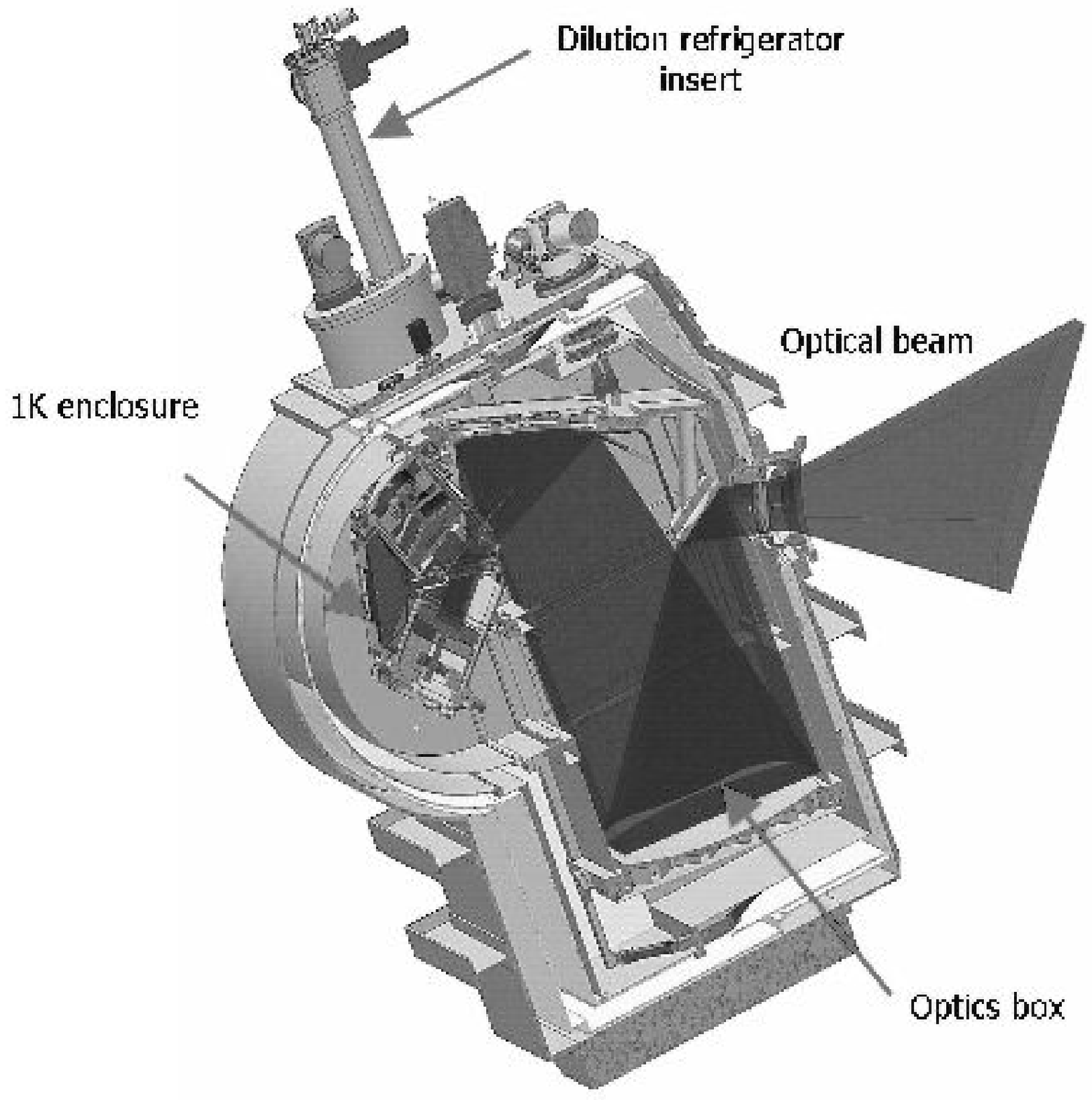}{393.120pt}{0}{77.7}{77.7}{-238.998pt}{-111.099pt}
\caption{The SCUBA-2 cryostat.  The cryostat is $\sim2.4\times2\times1.5\rm\ m$ in size and weighs 2.5~tonnes.}
\label{fig:cryostat}
\end{figure}

To achieve sky-background limited performance the arrays are operated at $\sim 120\rm\ mK$. The cryostat is designed to operate without 
liquid helium to reduce operational costs and to allow unattended operation of the instrument.  A dilution 
refrigerator will cool the detector stage to 60~mK and the focal plane unit to 1~K.  Superinsulation inside the vacuum vessel and a radiation shield at $\sim60\rm\ K$ will reduce the radiative load on the optics box.
Pulse tube coolers will cool the internal filters and optics at the 60~K (radiation shield) and 
8~K (optics box) stages. 
The two focal plane units are contained in an enclosure known as the ``$1\rm\ K$ box''.  The 1~K box and its design is described elsewhere in these proceedings\cite{Wood04}.  The 1~K box will be cooled from the still of the dilution refrigerator.   
The cryostat is shown in \Figure{fig:cryostat} and is described in detail elsewhere\cite{Gost04}.

\subsubsection{Cooling}
SCUBA-2 presents a challenge in cooling.  Not only do the detectors need to have a bath temperature well below $100\rm\ mK$, but the cold optics, weighing almost $200\rm\ kg$ need to be cooled down to and maintained at $\sim8\rm\ K$.  To keep costs down, a liquid-helium free system is required.  A liquid nitrogen pre-cool system will help cool the cryostat from 300 to $77\rm K$.  Two Cryomech PT410 pulse tube coolers operating in parallel will cool the radiation shield and the optics box.  Running at $60\rm\ Hz$, the two pulse tube coolers will provide a total cooling power of $40\rm\ W$ at $45\rm\ K$ on the radiation shield and $1.0\rm\ W$ at $4.2\rm\ K$ on the second stage.  The pulse tube coolers will be connected to the radiation shield and optics box by flexible wicks which provide mechanical isolation.

SCUBA-2 will use a commercially-sourced, high-power, cryogen-free dilution refrigerator with a cooling power of $\sim 20\rm\ \mu W$ at $25\rm\ mK$.  It will use oil-free pumps to avoid blockages in the capilliaries that have been a problem with SCUBA.  The design of this refrigerator, based on a Leiden Cryogenics MNK125-500 dilution refrigerator, is well advanced.  A third PT410 cooler will pre-cool the $^3$He flowing into the dilution refrigerator.  The PT410 coolers have lower vibration than GM coolers, which means they are less likely to heat the mixing chamber mechanically.

\subsection{Magnetic Shielding}
SCUBA-2 uses SQUIDs for multiplexing and amplifying signals.  SQUIDs are extremely sensitive magnetometers.  This means that they must be well shielded from external magnetic fields.  Our strategy is to shield the detectors by incorporating a superconducting coating on the 1~K box and a superconducting film behind the detectors. This strategy depends on the Meissner effect to exclude magnetic fields from the interior of a superconductor.  

Shielding is most important for the SQUID series array amplifiers since these are the most sensitive devices.  Eight SQUID array amplifiers are packaged in a rectangular niobium box that surrounds a mu-metal sleeve. These enclosures have been proven effective.  The rationale behind this design is that the niobium will exclude magnetic flux and the mu-metal will attenuate any flux trapped in the niobium.

The field near the arrays must be kept below 100~nT at frequencies we care about (0--200~Hz).  This will be accomplished by covering the 1~K box with niobium foil.  High-permeability shields on the inside of the vacuum vessel will reduce the field at the 1~K box to less than one tenth of the earth's field to avoid trapping flux in the niobium.  We are required to leave a large hole (the cold stop) in the 1~K box to let sub-millimetre radiation in.  A layer of high-permeability material under the niobium will provide a low reluctance path for flux that enters through the cold stop and reduce its effect on the detectors.  
If we assume, that the magnetic field in the cryostat is equal to one tenth of the earth's magnetic field ($5\rm\ \mu T$) the 1~K box must have a shielding factor of $\sim50$ in order to meet the 100~nT requirement.  The results of magnetic modelling suggest that this will be easily achievable.  The exact amount of shielding required to keep the field below $5\rm\ \mu T$ at the 1~K box depends on the details of the magnetic environment at the telescope.  If necessary, mu-metal shields can be attached to the outside of the cryostat to reduce further the field inside.  The magnetic environment at the telescope will be surveyed to see if this is necessary.

Sources of magnetic fields inside the cryostat (e.g. from the motor that operates the shutter; see \Section{sect:DREAM}) will be evaluated using finite-element magnetic modeling and, where necessary, will be shielded with high-permeability and/or superconducting foil.  The shutter motor will be configured in such a way as to have a low magnetic moment and will be surrounded by high-permeability magnetic shielding.
There will be a $150\rm\ nm$ layer of molybdenum on the back of the MUX wafers.  This will act as a superconducting ground plane behind the arrays to shield them from fields generated by Johnson noise currents in the beryllium copper hairbrush.  It will also help shield against external fields.

\subsection{Multi-Channel Electronics}
\label{sect:electronics}
Woven ribbon cables bring the signals from the SQUID series arrays to a readout card 
incorporating an FPGA.  Output values are co-added and stored in the 
FPGA before being sent over a fibre optic link to a data collection PC 
(one per sub-array) at a rate of 200 frames per second.  This frame rate has been chosen to allow fast scanning of the telescope over the sky for large-scale surveys (see \Section{sect:surveys}).  Each frame samples half a beam in fast scanning mode.
The multi-channel electronics (MCE) is currently undergoing testing and firmware development is progressing.  It is expected that the MCE will be integrated into the sub-array prototype test-bed in September 2004.

\subsection{Optics}
The field of view of the JCMT has a diameter of 600~mm at the Cassegrain focus.  It would be impractical to fill such a large area with a cryogenic detector array.  Thus, the focal plane has to be re-imaged to match an array size of 100~mm square.  The receiver cabin that rotates with the telescope is too small to accomodate the optics needed so the beam is directed through a narrow bearing tube to the mezzanine level where the SCUBA-2 instrument sits.  The aluminium mirrors for SCUBA-2's optics have been designed and are being manufactured.  The cold mirrors which go in the cryostat have already been delivered.

\subsection{Observing Modes}
\subsubsection{Survey Modes}
\label{sect:surveys}
One of SCUBA-2's main scientific goals is to carry out wide-field surveys.  Large-scale surveys will be carried out by scanning the telescope at speeds up to 600~arsec/sec.  It is intended that these surveys will be carried out semi-automatically, possibly remotely.
\subsubsection{Imaging Mode}
\label{sect:DREAM}
SCUBA-2 has dc-coupled pixels which will be used for deep imaging to the confusion limit of an $8\times8$~arcmin area of sky.  The flat-field accuracy required  for a one-hour observation is estimated to be one part in $10^7$.
The dark SQUIDs will mitigate drift downstream of the detectors.
Flat-fielding will be done continuously by dithering the SMU in DREAM mode\cite{lePo98}.  In this scheme each pixel makes a mini-map which overlaps those of its neighbours.  The results are combined to subtract the sky background.
A cold shutter at the entrance to the $1\rm\ K$ box will take dark frames to compensate for any drifts in the detectors and read out electronics. 
The shutter blade will be heatsunk to the 1~K box by a flexible heat strap, while the motor will be heatsunk to the optics box.

\subsubsection{Polarimetry}
SCUBA-2 will be equipped with a polarimeter mounted in front of the cryostat window\cite{Bast04}.  This will be a scaled up version of the SCUBA polarimeter\cite{Grea03} with the modification that the waveplate will rotate continuously.  The waveplate is currently being designed.  The polarimeter will allow us to map magnetic fields in Galactic star-forming clouds\cite{Matt04} and measure the synchrotron polarization of black hole candidates.

\subsubsection{Spectroscopy}
We are designing an imaging Fourier transform spectrometer (FTS) based on the Mach-Zender design\cite{SPIE02FTS}.  The FTS will cover at least a quarter of the field of view of the JCMT and will be used primarily for imaging spectroscopy of extended Galactic sources but will also provide useful information on bright nearby galaxies and planetary atmospheres.

\section{CURRENT STATUS}
SCUBA-2 is due to be delivered to the JCMT in February 2006.  
The first prototype sub-array has been manufactured and is being integrated for testing.  
The optics and cryostat  are being manufactured.
We expect to begin testing the prototype sub-array shortly and hope to know whether we are ready to manufacture science-grade arrays by September 2004.  Delivery to the telescope is scheduled for February 2006.

\acknowledgments     
SCUBA-2 is a collaboration between the UK Astronomy Technology Centre (Edinburgh), the National Institute for Standards and Technology (Boulder), the Scottish Microelectronics Centre (Edinburgh), the University of Wales (Cardiff), the Joint Astronomy Centre (Hawaii), the University of Waterloo, the University of British Columbia (Vancouver), the University of Lethbridge, Saint Mary's University (Halifax), and Universit\' e de Montr\' eal.  The project is funded by the UK Particle Physics and Astronomy Research Council, the JCMT Development
Fund and the Canada Foundation for Innovation.


\bibliography{SCUBA-2}   
\bibliographystyle{spiebib}   

\end{document}